# FABRICATION OF MEM RESONATORS IN THIN SOI

*Daniel GROGG, Nicoleta Diana BADILA-CIRESSAN, Adrian Mihai IONESCU*

Micro/Nano Electronics Laboratory, Ecole Polytechnique Fédérale de Lausanne (EPFL), Switzerland
{daniel.grogg, diana.ciressan, adrian.ionescu}@epfl.ch

**ABSTRACT**

A simple and fast process for micro-electromechanical (MEM) resonators with deep sub-micron transduction gaps in thin SOI is presented in this paper. Thin SOI wafers are important for advanced CMOS technology and thus are evaluated as resonator substrates for future co-integration with CMOS circuitry on a single chip. As the transduction capacitance scales with the resonator thickness, it is important to fabricate deep sub-micron trenches in order to achieve a good capacitive coupling. Through the combination of conventional UV-lithography and focused ion beam (FIB) milling the process needs only two lithography steps, enabling therefore a way for fast prototyping of MEM-resonators. Different FIB parameters and etching parameters are compared in this paper and their effect on the process are reported.

## 1. INTRODUCTION

Silicon micro-electromechanical (MEM) resonators have shown performances comparable to those of quartz resonators [1, 2] but they offer higher levels of integration in electronic circuits. The research effort for these devices is currently focused on rather thick layers combined with deep sub-micron gaps (also referred to as nanogaps), as this allows reduction of the resonator's equivalent resistance, which is given for small amplitudes [1].

$$R_{eqv} = \frac{\sqrt{km}}{QV_P^2 \varepsilon_0^2} \frac{g^4}{h^2 w^2}$$

In the above statement, k and m are the effective spring constant and the effective mass, Q is the resonators quality factor, $V_P$ is the applied polarization voltage and $\varepsilon_0$ is the free space permittivity. The dimensions of the transduction gap are the height (h) and the width (w) of the electrode and g is the gap dimension. Different fabrication processes have been proposed to create nanogap resonators in thick SOI [3, 4], achieving good resonator performance. To further pursue the integration of MEMS and CMOS in the perspective of the decreasing SOI thickness currently used with state of the art CMOS technologies, we evaluated the fabrication of MEMS resonators in thin SOI. This type of substrate seems extremely promising to profit from the most advanced CMOS and MEMS technology at the same time. A fabrication process is proposed to produce single crystalline silicon MEMS resonators with gaps as smaller than 100nm.

## 2. DEVICE DESCRIPTION AND FABRICATION

### 2.1. Design

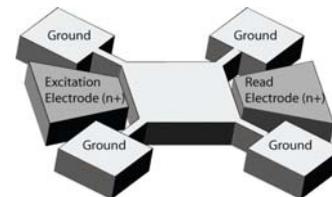

Figure 1: Schematic view of a resonator with capacitive detection.

Square resonators with support in all four corners have been designed, for higher mechanical stability. The energy conversion between the electrical and the mechanical domain is done with electrostatic transduction gaps. Different designs are tested with either 2 or 3 electrodes. The gap dimension is important, as it influences directly the motional resistance and the DC bias needed to excite the resonator. The resonators are designed to vibrate in a Lamé–mode and the resonance frequencies simulated with ANSYS® are given in Table 1. With the given electrode configuration, it is also possible to excite a square extensional mode, where all four sides are moving in phase, but such a mode has no nodal point on the circumference and its performance is not optimized. To simplify the FIB milling step, all designs were composed only of straight lines, as our FIB equipment did not allow to create rounded shapes easily.

| Resonator size | Resonance Frequency |
|---|---|
| 10 µm | 318 MHz |
| 20 µm | 162 MHz |
| 30 µm | 117 MHz |
| 40 µm | 87 MHz |
| 50 µm | 70 MHz |
| Table 1: Resonance Frequencies simulated with ANSYS® | |

### 2.2. Processflow

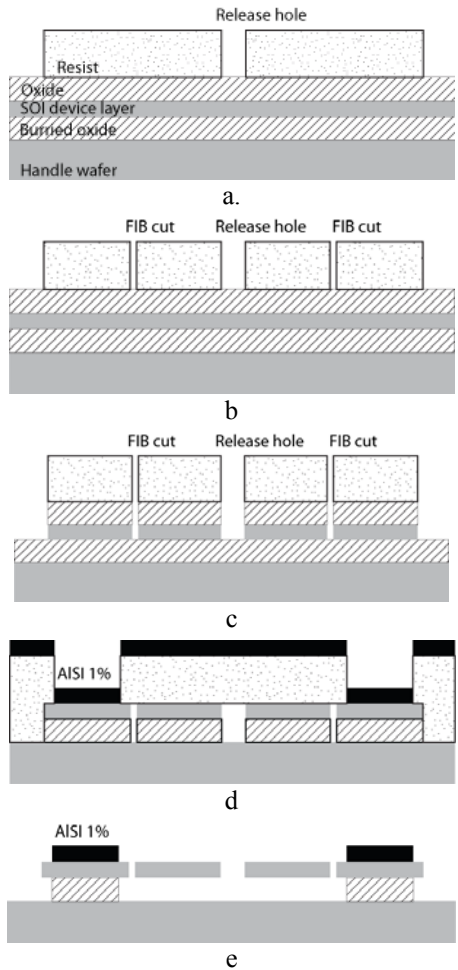

Figure 2: Schematic of the process to create MEMS resonators with only two levels of photolithography.

Thin SOI wafers with a buried oxide thickness of 400 nm and a SOI device layer of 340 nm were used. The SOI wafer is heavily doped with Phosphor to achieve good electrical contact. Then a thermal oxide is grown on top, which will serve as stop layer for the FIB milling and as hardmask for the silicon etching.

The fabrication is based upon a combination of standard UV lithography with a resolution of 800 nm and subsequent FIB milling of the photoresist to fabricate single crystalline silicon resonators and electrodes with a single mask step. A 500 nm thin photoresist (Shipley, S1805) was coated on the wafer and patterned by UV lithography (Figure 1a). The resist was then locally milled away to create gaps in the order of 50 nm to 100 nm (Figure 2b and Figure 3).

As it can be seen in Figure 3a, the FIB milling has been optimized to stop the Ga+-Ions in the silicon dioxide layer and avoid Ga-contamination of the SOI device layer. This is important in order to avoid local masking effects that can occur during the reactive ion etching and to keep the option for additional electrical functionality.

Reactive ion etching was used to transfer the patterns created by UV-lithography and FIB into the silicon dioxide hard mask during the same etch step. In the following silicon etch (Figure 1c and Figure 4), two different etch processes (chlorine or fluorine chemistry) were compared with respect to their possible influence on the gap size. After the patterning of the silicon, the silicon dioxide hardmask was removed with a short BHF 7:1 etch. AlSi 1% was sputter deposited onto a two-layer lift-off resist (Microchem LOR5A; Shipley S1813) to create the contacts on the pads (Figure 1d). The metal layer was annealed at 425°C. Finally the resonators are released using silox (pad etch) to etch the buried oxide under the resonator away (Figure 1e) without etching the metal layer, followed by a $CO_2$ supercritical drying step.

### 3. FOCUSSED ION BEAM MILLING

The combination of the appropriate FIB milling parameters together with a suitable etch process is a key element for rapid prototyping of nanostructures. This approach is interesting for validation of new nanogaps MEMS devices, but has its limitations for high throughput applications. We tested Ga+-Ion milling with and without $H_2O$ enhancement [5]. The tests have been performed on a FEI Nova 600 Nanolab, a SEM / FIB dualbeam systems which is also equipped with a gas injection system (GIS).

Preliminary tests were done to determine ideal tradeoff between beam current and resulting size of the nanogaps in the resist. The theoretical minimal beam diameter is depending on the beam current and a smaller current is therefore promising a narrower gap, but a small current is also increasing the milling time and therefore slowing down the whole process. For our experiments, a tradeoff between time and resolution has been found with an ion current of 50 pA at 30 keV acceleration voltage (theoretical beam diameter of 19 nm) and used throughout all experiments.

Once the milling current was fixed, the milling time was adjusted so to stop precisely in the silicon dioxide layer to avoid Ga contamination of the silicon layer. On the FEI dualbeam machine interface, time is automatically calculated as a function of surface and depth (milling volume), therefore we indicate all timescales as an equivalent depth for a single line (beam overlap is 50 %, dwelling time is 1 µs).

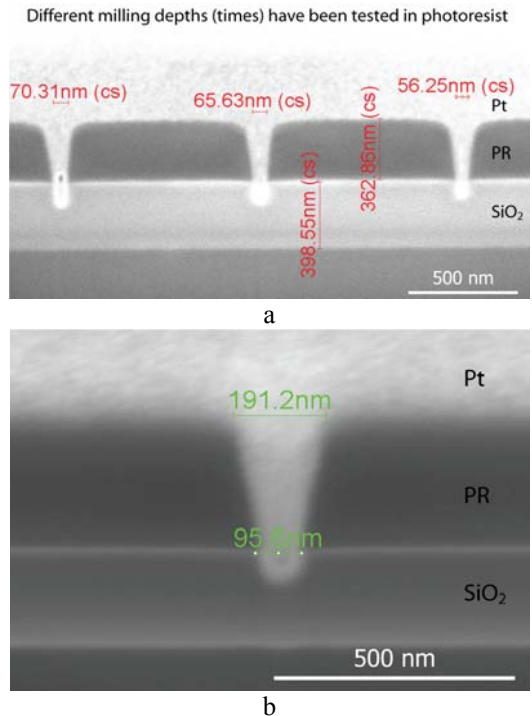

Figure 3: Cross-section of a FIB milling test on SiO2 layer. (a) a 400 nm thick SOI layer was used to control the milling depth of Ga+ FIB in the resist without H2O. (b) The same test has been done on 300 nm thick SOI using H2O enhanced FIB milling.

SRIM (http://www.srim.org/) simulation of Ga+ ions in a silicon-dioxide layer results in a stopping range of 274 A and shows that all ions are stopped within approximately 600 A. It seems therefore safe to stop the ion milling within the first 100 nm, to assure Ga free silicon device layer. The result of a milling test for equivalent depth of 700, 800 and 900 nm (in silicon) are shown in Figure 3a, where 800 nm was chosen to be suitable for the process.

### 3.1. FIB milling with H2O

The GIS enables the use of different gases for either deposition or etching. A selective carbon milling (Magnesium Sulfate Heptahydrate) process is compared to the conventional milling technique. The effect of the selective carbon milling seems to depend upon the product of the sputtered material and $H_2O$. The result is an increased milling efficiency of Ga+-ions on carbon containing materials (thus polymers also) and at the same time a reduction of the milling efficiency on $SiO_2$. With this technique and an ion beam current of 50 pA, the equivalent depth could be reduced to 500 nm. It has been noted that the aspect ratio of the $H_2O$ enhanced milling is strongly reduced, as can be seen in Figure 3b.

### 3.1. Charging and edge effects

Charging has been an additional difficulty in this process, as the Ga+-ion beam is drifting away from its initial position during the milling. This drift is time and current dependent, but it could effectively be canceled with a strong electron current from a separate electron gun.

Two types of problems occurred at the photoresist edge. The photoresist sidewall has an angle of approximately 45° (in a cross-section), making it difficult to align the nanogaps precisely. Small overlaps of the designed gap with the sidewall were leading to shortcuts between the resonator and the electrodes. To reduce this problem, small circles or triangles were added on the resist edge. The second effect also appeared on the edge of the resist along the milled nanogaps. As can be seen on the figure 5b, the gap is thinning towards the end, but this effect was also reduced, although not completely avoided, with the added structures.

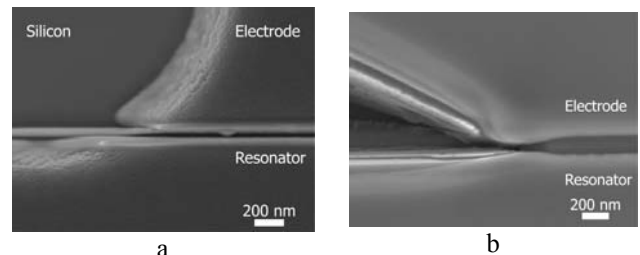

Figure 4: Topview of a nanogaps after silicon dioxide etching without (a) and with an added triangle (b).

These two problems did not occur with the $H_2O$ enhanced FIB milling. In fact, due to the reduced thickness at the edge of the resist, this type of milling resulted in a slightly increased gap, leading to a smooth transition between the gap and the open area.

### 3. NANOGAP ETCHING

The transfer of the photoresist mask into the silicon dioxide was done using $SF_6$ based $SiO_2$ dry etching process, with a selectivity of 1:1 and an etch speed of 280 nm/min. This etch step did not show any dependence on the Ga ion concentration in the $SiO_2$ layer.

Two etch processes were compared to qualify the transfer of the $SiO_2$ hard mask into the silicon. A fluorine ($SF_6$) based silicon RIE etch (Si etch rate of 1 µm/min, selectivity 20:1 on $SiO_2$) was compared to chlorine ($Cl_2$) based ICP etch (Si etch rate of 0.35 μm/min, selectivity 1:1 on $SiO_2$). The combination of conventional FIB milling and high selectivity fluorine etch process resulted in very small gaps, but a subsequent cross-section revealed, that this process suffers from a rather large notching problem (Figure 4a). The gap is much wider than expected over half of the resonator layer.

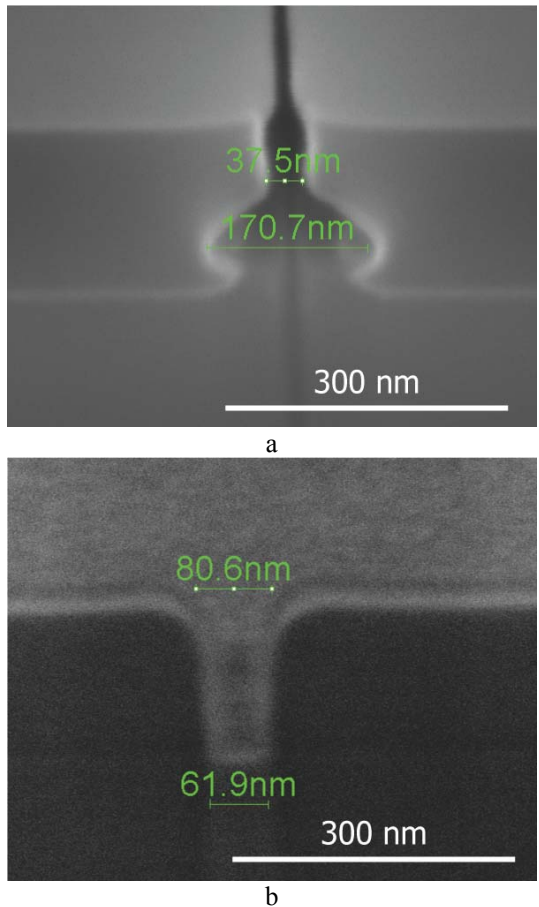

Figure 5: Nanogaps in thin SOI created with (a) FIB without enhancement and fluorine etching (a) $H_2O$ enhanced FIB and chlorine etching and

The chlorine etching process was successfully used to transfer the larger gaps, created by $H_2O$ enhanced FIB, into to the SOI device layer. A cross section showing the gap profile is shown in Figure 4b. Gaps with widths of clearly less than 100nm and almost vertical sidewalls have been created with this process. The resulting thin SOI MEMS resonator with nanogaps is shown in Figure 6.

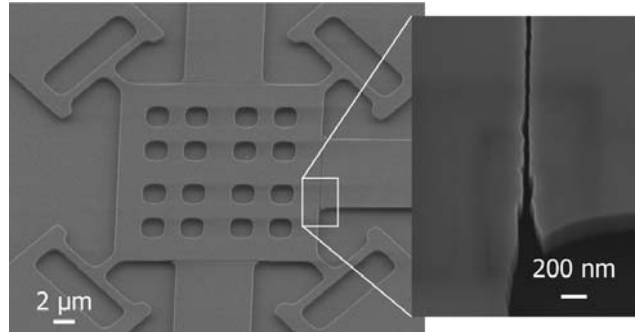

Figure 6: Bulk lateral resonator with side length L = 20 μm (designed freq: 162 MHz) and nano-gap (< 100 nm) detection fabricated with the proposed FIB prototyping process.

### 4. CONCLUSION

We have successfully demonstrated a rapid FIB prototyping process for fabricating deep sub-micron gap MEMS resonators in thin SOI. Only two mask levels are used. The impact of 2 different FIB milling processes combined with two different silicon etches have been investigated, to optimize the fabrication of sub 100nm gaps in thin silicon. The first thin film bulk lateral resonators in thin SOI have been created.